\documentclass[reprint,showkeys,amsmath,amssymb,aps,pra,floatfix]{revtex4-1}

\pdfoutput=1
\usepackage[english]{babel}
\usepackage{amsmath}
\usepackage{amssymb}
\usepackage[caption=false]{subfig}
\usepackage{hyperref}
\usepackage{graphicx}
\usepackage{xcolor}
\usepackage{floatrow}
\DeclareCaptionLabelFormat{r-parens}{\textbf{#2}}
\makeatletter
\pdfpageheight\paperheight
\pdfpagewidth\paperwidth
\makeatother
\newcommand{\tTKT}{\tilde T_{\rm \scriptscriptstyle KT}}
\newcommand{\TKT}{T_{\rm \scriptscriptstyle KT}}

\newcommand{\W}{\vee\!\vee}
\newcommand{\badW}{\,\!^{-\!}\vee}

\begin{document}
\title{Machine learning vortices at the Kosterlitz-Thouless transition}
\author{Matthew J.~S.~Beach} \email{matthew.beach@uwaterloo.ca}
\author{Anna Golubeva}
\author{Roger G.~Melko}
\affiliation{Department of Physics and Astronomy, University of Waterloo, Waterloo N2L 3G1, Canada}
\affiliation{Perimeter Institute for Theoretical Physics, Waterloo, Ontario N2L 2Y5, Canada}
\date{\today}

\begin{abstract}
	Efficient and automated classification of phases from minimally processed data is one goal of machine learning in condensed matter and statistical physics. Supervised algorithms trained on raw samples of microstates can successfully detect conventional phase transitions via learning a bulk feature such as an order parameter. In this paper, we investigate whether neural networks can learn to classify phases based on topological defects. We address this question on the two-dimensional classical XY model which exhibits a Kosterlitz-Thouless transition. We find significant feature engineering of the raw spin states is required to convincingly claim that features of the vortex configurations are responsible for learning the transition temperature. We further show a single-layer network does not correctly classify the phases of the XY model, while a convolutional network easily performs classification by learning the global magnetization. Finally, we design a deep network capable of learning vortices without feature engineering. We demonstrate the detection of vortices does not necessarily result in the best classification accuracy, especially for lattices of less than approximately 1000 spins. For larger systems, it remains a difficult task to learn vortices.
\end{abstract}
\maketitle

\section{\label{sec:intro} Introduction}
	The remarkable success of artificial neural networks in the tasks of image recognition and natural language processing has prompted interdisciplinary efforts to investigate how these new tools might benefit a broad range of sciences. One of the most intriguing areas of application is condensed matter physics, where the exponentially large Hilbert space of a quantum many-body state provides the ultimate big data set. In fields such as computer vision, it has been demonstrated that neural networks have the ability to extract physical features from highly complex datasets~\cite{zeiler_visualizing_2014,lecun_deep_2015,guo_deep_2016,krizhevsky_imagenet_2017}. This gives hope that machine learning techniques may provide a tool to probe regions of the many-body Hilbert space that are currently intractable with other algorithms.

	In the realm of classical statistical physics, supervised and unsupervised learning have been applied successfully to classify symmetry-broken phases~\cite{carrasquilla_machine_2017,wetzel_machine_2017,van_nieuwenburg_learning_2017,ponte_kernel_2017}. In some cases, it is possible to deduce that the network has learned an order parameter or another thermodynamic quantity~\cite{carrasquilla_machine_2017,wetzel_machine_2017,ponte_kernel_2017}. This interpretability is one major advantage of data sets derived from statistical physics, and can contribute to the theoretical understanding of the behavior of neural networks in real-world applications.

	Motivated by the successful application of supervised learning to conventional symmetry-breaking transitions, it is natural to ask whether neural networks are capable of distinguishing unconventional phase transitions driven by the emergence of topological defects. The prototypical example for such a system is the two-dimensional XY model, which exhibits a Kosterlitz-Thouless (KT) transition~\cite{kosterlitz_ordering_1973}. Several unsupervised learning strategies have been applied to this model previously, for example, it was found that principle component analysis (PCA)~\cite{jolliffe_principal_2002} performed on spin configurations captures the magnetization which is present in finite-size lattices~\cite{wetzel_unsupervised_2017,hu_discovering_2017,wang_unsupervised_2017}. Even when trained directly on vorticity, PCA is unable to resolve vortex-antivortex unbinding, which is attributed to the linearity of this method~\cite{hu_discovering_2017}. Similarly, variational auto-encoders~\cite{kingma_auto-encoding_2013}, a popular tool for unsupervised learning based on Bayesian inference, perform classification by learning a bulk magnetization~\cite{wetzel_unsupervised_2017,cristoforetti_towards_2017,wang_unsupervised_2017}.

	 In contrast, efforts in supervised learning have been more successful, although none have been applied directly to the XY model. In Ref.~\cite{broecker_quantum_2017}, a convolutional network trained on winding numbers correctly classified interacting boson phases separated by a KT transition. However, this same method failed when trained directly on raw configurations. A related problem was explored in Ref.~\cite{zhang_machine_2017}, where the authors trained a convolutional network directly on Hamiltonians of one-dimensional topological band insulators labeled by their global winding number. By inspecting the trained weights the authors deduced that the network had learned to calculate the winding number correctly.

	In this paper, we apply several supervised machine learning strategies to the task of identifying the KT transition in the two-dimensional XY model. We ask whether it is possible for a neural network, trained only on raw spin states labeled by their phases, to learn a latent representation that can be interpreted as the local vorticity of the spin variables. First, we compare supervised learning algorithms involving feed-forward and convolutional neural networks applied to both unprocessed (raw spin configurations) as well as processed input data (vorticity). We then use both types of input data in the semi-supervised confusion scheme from Ref.~\cite{van_nieuwenburg_learning_2017}. Lastly, we explore to which degree feature engineering of the raw spin configurations is required, and whether the network can learn to process the data into something resembling vortices using additional convolutional layers.

	Although is it possible to learn vortices, the network can also perform its classification task to reasonable accuracy by finding a local optimization minimum which is unrelated to topological features. We conclude with a discussion on the challenging task of seeing the vortex-antivortex unbinding transition in the two-dimensional XY model using machine learning techniques.

\section{\label{sec:back} Background}

	The classical XY model consists of unit spins with nearest-neighbor interactions given by
	\begin{equation}
		\mathcal{H}_{XY} = -J \sum_{\langle ij \rangle} \cos \left(\theta_i - \theta_j\right),
	\end{equation}
	where $\langle ij \rangle$ indicates that the sum is taken over nearest neighbors and the angle $\theta_i \in [0, 2\pi )$ denotes the spin orientation on site~$i$. Although the Mermin-Wagner theorem states that a long-range ordered (LRO) phase cannot exist in two dimensions due to the coherence of massless spin waves~\cite{mermin_absence_1966}, the formation of topological defects (i.e., vortices/antivortices) in the XY model results in a quasi-LRO phase~\cite{kosterlitz_ordering_1973,kosterlitz_critical_1974}. The transition between the low-temperature quasi-LRO phase with an algebraically decaying correlation function and the high-temperature disordered phase with an exponentially decaying correlation function is a KT transition and the associated temperature is denoted as $\TKT$. Transitions of this universality class can be found in a variety of systems, with one of the most famous being the superfluid transition in two-dimensional helium~\cite{kapitza_viscosity_1938,misener_flow_1938,bishop_reppy_1978}.

\begin{figure}
		\centering	
		\fbox{\hspace{0.25cm}\includegraphics{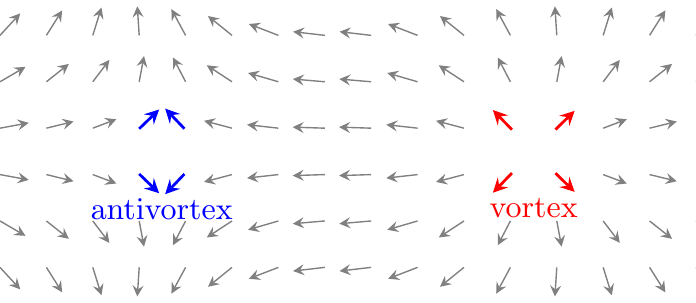}}
		\caption{\label{fig:fig1}
		A example of a vortex and antivortex in the XY model on the lattice. A vortex has winding number $k=1$, while an antivortex has $k=-1$. }
	\end{figure}

	The topological defects in the XY model are quantified through the vorticity $v$, defined as
	\begin{equation}
		\label{eq:vorticity}
		v \equiv \oint_C \nabla \theta \cdot d \vec \ell = 2\pi k, \qquad k= \pm1, \pm2,... \,,
	\end{equation}	
	where $C$ denotes any closed path around the vortex core and $k$ is the winding number of the associated spins. A vortex is defined by positive winding number, $k=1$, and an antivortex by $k=-1$. On a lattice, the integral may be approximated by the sum of the angle differences over a plaquette. An example of a vortex and antivortex is shown in Fig.~\ref{fig:fig1}.

\begin{figure*}
	 	\sidesubfloat[]{\hspace{-0.37cm}\includegraphics{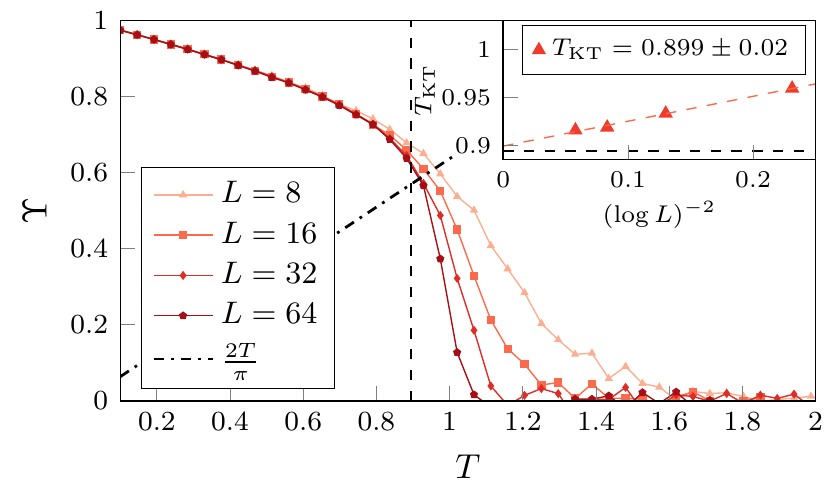}
	 	\label{fig:fig2a}} \,\,\,
	 	\sidesubfloat[]{\hspace{-0.37cm}\includegraphics{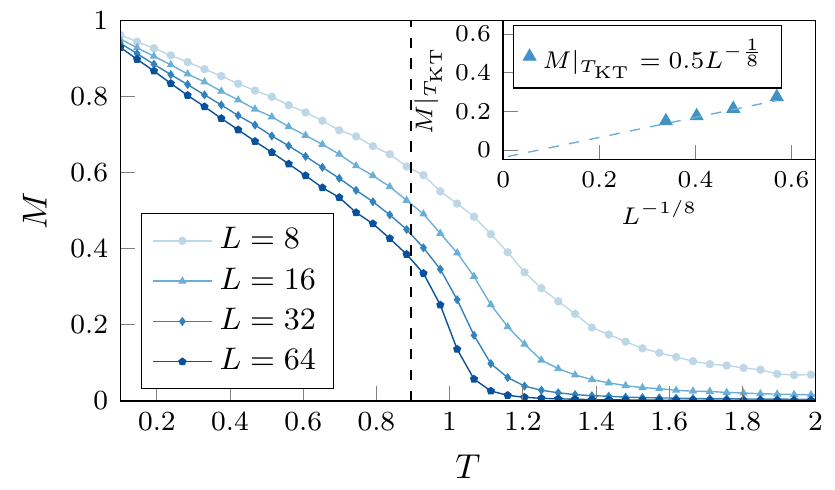}
	 	\label{fig:fig2b}}
		\caption{\label{fig:fig2}
		Estimators of the XY model on a $L\times L$ lattice with periodic boundary conditions computed via Monte Carlo sampling.
		{\bf (a)} shows the helicity modulus for various lattice sizes $L$. The estimated critical point $\tTKT$ is determined by the Nelson-Kosterlitz universal jump where the helicity modulus, $\Upsilon$, intersects the line $\frac{2 T}{\pi}$. The inset shows how $\tTKT$ scales with $(\log L)^{-2}$ towards the thermodynamic $\TKT$ shown by the black dashed line.
		{\bf (b)} shows the non-zero magnetization present in the finite-size XY model. The magnetization vanishes as $L^{-\frac{1}{8}}$ in the thermodynamic limit with the scaling shown in the inset. }
	\end{figure*}

	Below $\TKT$, vortex-antivortex pairs form due to thermal fluctuations, but they remain bound to minimize their total free energy. At $\TKT$, the entropy contribution to the free energy equals the binding energy of a pair, triggering vortex unbinding which drives the KT phase transition. The essential singularity of the free energy at $\TKT$ means that all derivatives are finite at the transition. For example, the specific heat is observed to be smooth at the transition, with a non-universal peak at a $T > \TKT$ which is associated with the entropy released when most vortex pairs unbind~\cite{chaikin_principles_2000}. While the thermodynamic limit of the XY model has strictly zero magnetization for all $T>0$, a non-zero value is found for systems of finite size (see Fig.~\ref{fig:fig2b})~\cite{chung_essential_1999,olsson_monte_1995}.

	One method to calculate $\TKT$ from finite-size data is to exploit the Nelson-Kosterlitz universal jump~\cite{nelson_universal_1977,minnhagen_superfluid_1981}. This is determined from where the helicity modulus, $\Upsilon$, crosses $\frac{2 T}{\pi}$. The helicity modulus, also called spin wave stiffness or spin rigidity, measures the response of a system to a twist in the boundary conditions (i.e., torsion). From the linearized renormalization group (RG) equations, one can derive the finite-size scaling behavior of the critical temperature $\tTKT$ on a $L \times L$ lattice to be
	\begin{equation}
		\label{eq:scaling}
		\tTKT (L) \approx \TKT + \frac{\pi^2}{4c(\log L)^2}\, ,
	\end{equation}
	with a constant $c$~\cite{nelson_universal_1977}. Fig.~\ref{fig:fig2a} shows the helicity modulus $\Upsilon$ and the scaling of $\TKT$ derived from Monte Carlo simulations. From our generated samples, we find $\TKT = 0.899 \pm 0.06$, which is consistent with the literature value of $\TKT=0.893$~\cite{olsson_monte_1995,chung_essential_1999,komura_large-scale_2012}. As shown in Fig.~\ref{fig:fig2b}, the magnetization evaluated at the critical point, $M|_{T_{\rm KT}}$, is of significant magnitude, and scales with $L^{-1/8}$ as expected~\cite{chung_essential_1999}, to within a $4\%$ error.

	In the next section, we explore which neural networks can accurately distinguish the phases above and below the thermodynamic temperature $\TKT$. We employ two standard network architectures motivated by canonical problems in machine learning (such as classification of the MNIST dataset) using XY spin configurations for finite-size systems as input data. Based on previous observations that conventional phase boundaries estimated by supervised learning follow established finite-size scaling~\cite{carrasquilla_machine_2017}, we compare the scaling of $\tTKT$ predicted by our neural networks with the $(\log L)^{-2}$ form above.

\section{Methods \& Results}

	We study the binary classification of the two phases of the XY model, labeling configurations as belonging to either the low $T<\TKT$ or high $T>\TKT$ temperature phases. Our goal is to confirm whether simple supervised learning with neural networks is capable of correctly classifying spin configurations according to these labels. In particular, we wish to interpret whether the network relies on the (finite-size) magnetization, or on topological defects. Further, we inquire as to what specific network architecture is required to achieve this goal and what features different architectures may utilize.

	We employ standard Monte Carlo simulation methods to generate spin configuration of the XY model~\cite{wolff_collective_1989,walter_introduction_2015}. For the training set, we generate 1000 configurations per temperature, with 64 temperatures ranging from 0.1 to 3.0, for lattice sizes $L=8, ..., 64$ in increments of $8$. The test set is generated separately, with 100 configurations per temperature. From the training data, we randomly select 10\% for cross-validation, in order to decrease the chance of overfitting and to identify a definitive stopping point for training using early stopping~\cite{wang_optimal_1995,prechelt_early_1997,prechelt_automatic_1998}.

	The network is trained to minimize the loss function $L(y^{\rm pred}, y^{\rm true})$, where $y^{\rm true}$ represents the true binary labels and $y^{\rm pred}$ the predicted ones. We take the loss function to be the standard cross-entropy
	\begin{equation}
		\label{eq:loss}
		L(y^{\rm pred}, y^{\rm true}) = -\sum_i y^{\rm true}_i \, \log y^{\rm pred}_i\,.
	\end{equation}
	The parameters of the network (weights and biases) are then optimized through back-propagation to minimize the loss function on the training data~\cite{lecun_deep_2015}. Each network is trained until the loss function \emph{evaluated on the validation set} fails to decrease after 50 training epochs. Early stopping with cross-validation is commonly used to choose the network parameters with minimal generalization error~\cite{prechelt_early_1997}.
	We implement the networks with the Keras library using the TensorFlow backend~\cite{chollet_keras_2015,abadi_tensorflow_2015}.

	We employ two different standard network architectures: a one-layer feed-forward network (FFNN) and a deep convolutional network (CNN). The FFNN consists of one hidden layer of 1024 sigmoid activation units and one sigmoid output unit. The CNN starts with a two-dimensional convolutional layer consisting of 8 filters of size $3\times 3$ with rectified linear (ReLu) activation functions. The output from this layer is passed to another identical convolution layer with 16 filters before applying $2\times 2$ max-pooling. The network is then reshaped and fed into a fully-connected layer with 32 ReLu units and passed to a single sigmoid output unit. Because there is a total of $1024\,L^2 + 2057$ trainable parameters in the FFNN, it can be difficult to train as compared to the $128\,L^2-1024\,L+3361$ parameters in the CNN. Further, the CNN explicitly takes advantage of the two-dimensional structure of the input to vastly improve performance, as we now discuss.

\subsection{\label{subsec:finite}Finite-size scaling of supervised learning}

	\begin{figure*}
	 	\sidesubfloat[]{\hspace{-.7cm}\includegraphics{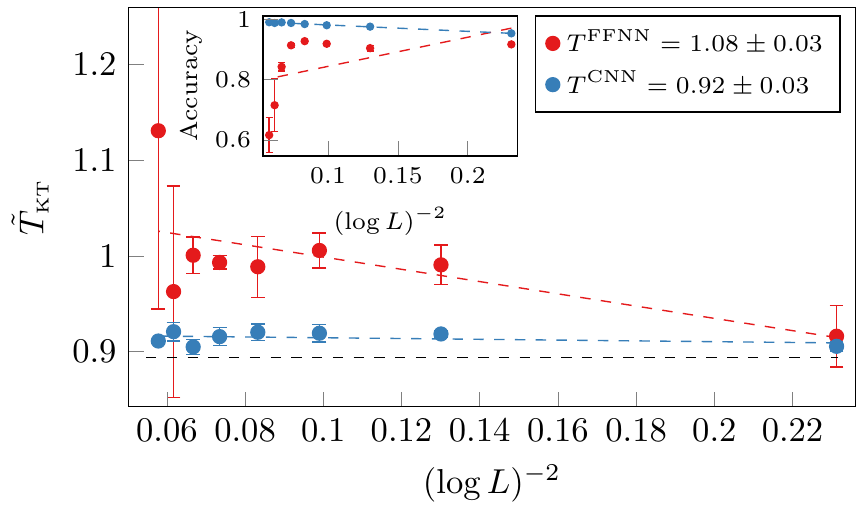}
	 	\label{fig:3a}} \,\,\,\,\,\,\,
	 	\sidesubfloat[]{\hspace{-.7cm}\includegraphics{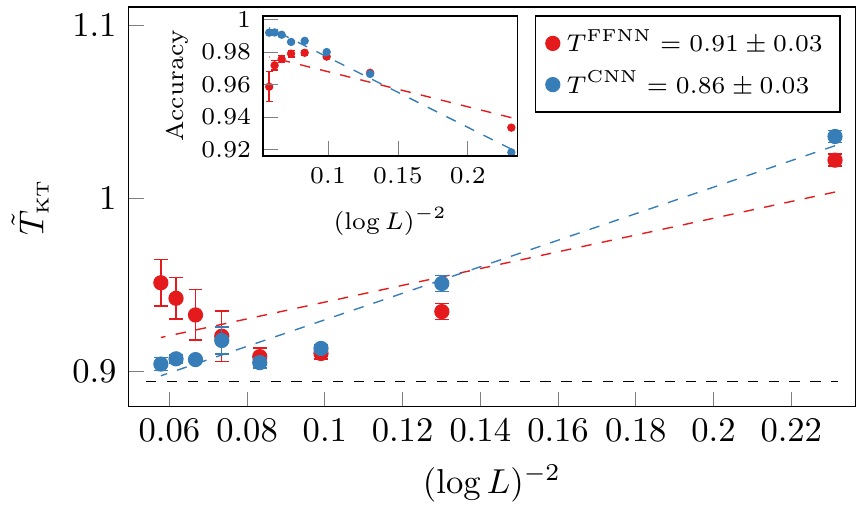}
	 	\label{fig:3b}}
		\caption{\label{fig:finite-scaling}
		Finite-size scaling of the predicted $\TKT$ for FFNN and CNN trained on either {\bf (a)} raw spin configurations, or {\bf (b)} the vorticity. In either case the FFNN performs worse than the CNN according to the test classification accuracy (insets). The critical temperature is determined by the point where the sigmoid output, as a function of temperature, crosses 0.5. Each data point and variance is obtained by training 10 networks with stochastic gradient descent until the validation loss function fails to improve after 50 epochs (early stopping).}
	\end{figure*}

	One goal in modern machine learning is to minimize the amount of feature engineering required. In our case, this corresponds to treating the raw spin configurations as direct inputs to the neural networks. For the XY model, this data is formatted as angle values, $\theta_i \in [0, 2\pi)$, on an $L \times L$ lattice with periodic boundary conditions.

	For a given sample configuration, the sigmoid output function gives the probability of the state belonging to the low- (or high-) temperature phase. The temperature of a configuration for which the output probability is exactly 0.5 can be interpreted as the critical temperature $\tTKT$. It is this point where the network is most uncertain about how to classify the input state. Interestingly, this point scales with the correct correlation length critical exponent and predicts the thermodynamic critical temperature accurately for the Ising model~\cite{carrasquilla_machine_2017}. In that case, training a FFNN with a single hidden layer of 100 sigmoid units was sufficient ($100\,L^2+202$ total parameters) to achieve high classification accuracy and correctly predict the critical temperature.

	Similarly, we study the performance of both a FFNN and a CNN in predicting $\TKT$ for the XY model. To get an estimate for the statistical variance, the training process is repeated ten times with different validation sets.

	As illustrated in the inset of Figure~\ref{fig:3a}, the FFNN has low classification accuracy (i.e., percentage of correctly classified configurations) for $L>48$. This results in the very poorly predicted critical temperature, $\tTKT$, in the main plot. In contrast, the accuracy of the CNN continually improves as $L$ increases. However, as evident from Fig.~\ref{fig:3a}, there is no clear finite-size scaling trend in the predicted $\TKT$. To interpret this we note that for each system size, the network is supervised on the thermodynamic value of $\TKT$. Thus, we speculate that each network could simply be learning to discriminate phases based on a robust, global feature which takes a unique value above and below $\TKT$ for any $L$.

	Based on previous experience, a global magnetization is a feature very easily detected in a supervised learning scheme~\cite{carrasquilla_machine_2017,wetzel_machine_2017,ponte_kernel_2017}. Since the finite-size configurations of the XY model themselves contain a non-zero magnetization at $T>0$ (see Fig.~\ref{fig:fig2b}), it is reasonable to hypothesize that the CNN simply learns this threshold value of the magnetization for each system size separately. Because of the Mermin-Wagner theorem, however, it is known that a global magnetization is not a relevant feature for $\TKT$ in the thermodynamic limit. Thus, in this case, some amount of feature engineering is crucial to achieve our goal of detecting a phase transition mediated by topological defects.

	In the next step, we preprocess the spin configurations into the associated vorticity and train the networks on these configurations. To calculate the vorticity, one computes the angle differences $\Delta\theta_{ij}\in [-2\pi, 2\pi]$ between each pair of neighboring spins $i$ and $j$ on a plaquette and converts these to the range $(-\pi, \pi]$. This can be done by applying the sawtooth function,
	\begin{equation}
	\label{eq:saw}
		\text{saw}(x) = \begin{cases}
		x + 2\pi, & x\leq -\pi, \\
		x, & -\pi \leq x \leq \pi, \\
		x - 2\pi, & \pi \leq x,
		\end{cases}
	\end{equation}
	to each $\Delta \theta_{ij}$. The sum of the rescaled angle differences gives the vorticity from Eq.~(\ref{eq:vorticity}).

	Trained on the vortex configurations, Fig.~\ref{fig:3b} shows that both the FFNN and CNN achieve high accuracy and scale with $L$ towards the correct value of $\TKT$. However, once again we observe that the FFNN begins to perform poorly for $L>32$, whereas the CNN continually improves. We note that the scaling seems consistent with Eq.~(\ref{eq:scaling}), particularly for the CNN. However, from this scaling alone, we cannot determine precisely what the CNN learns. For example, it could potentially classify the phases based on the sum of the squared vorticity (which is approximately zero below $\TKT$), or it might represent a more complicated function such as the average distance between vortex-antivortex pairs. Regardless, the scaling behavior may serve as a useful diagnostic to determine whether a given network is learning bulk features or topological effects.

\subsection{Learning by confusion}
	\begin{figure*}
	 	\sidesubfloat[]{\hspace{-0.6cm}\includegraphics{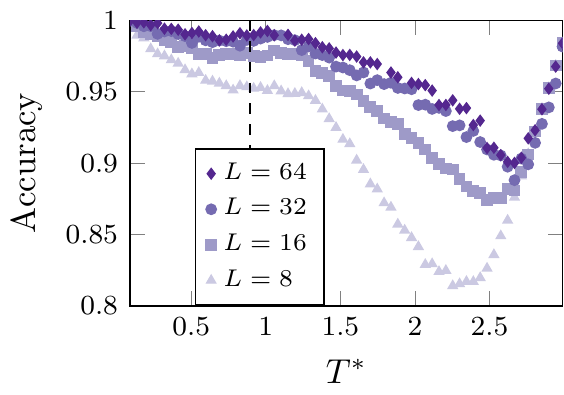}\label{fig:4a}}\,
	 	\sidesubfloat[]{\hspace{-0.3cm}\includegraphics{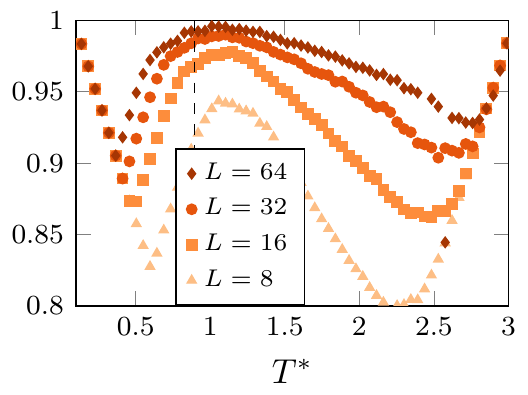}\label{fig:4b}}\,
	 	\sidesubfloat[]{\hspace{-0.3cm}\includegraphics{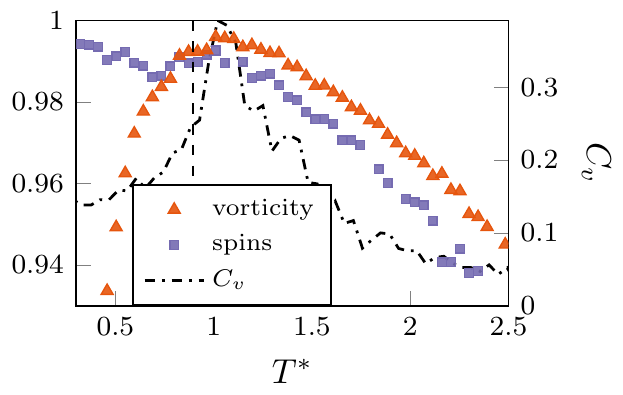}\label{fig:4c}}
		\caption{\label{fig:fig4}
		The learning by confusion scheme for a CNN applied to: {\bf (a)} raw spin configurations, {\bf (b)} vorticity configurations. The test accuracy is expected to form a $\W$ shape with the center peak at $T^*=\TKT$. In {\bf (c)}, the peak in specific heat ($C_v$) is compared to the peak of the test accuracy for a system of size $L=64$. The dashed vertical line shows the thermodynamic $\TKT$.}
	\end{figure*}

	We further investigate the difference between training on spin configurations and vortex configurations by employing a confusion scheme~\cite{van_nieuwenburg_learning_2017,liu_self-learning_2017}. Learning by confusion offers a semi-supervised approach to finding the critical temperature separating two phases by training many supervised networks on data that is deliberately mislabeled. The binary label `$0$' is assigned to a configuration if its temperature is less than a proposed $T^*$ and `$1$' otherwise. A new network is trained on each new labeling of the data, (i.e., for each $T^*$). It is expected that the highest accuracy is achieved when the labeling is close to the true value, and, trivially, at the end points. This results in a $\W$ shape when plotting the test accuracy as a function of $T^*$~\cite{van_nieuwenburg_learning_2017}. The peak on either endpoint can be attributed to the network being trained and tested exclusively on one class, in which case it will always place test data into that class. The key assumption in the confusion scheme is the existence of a true physical labeling of the data which the network is capable of learning more accurately than false labellings.

	Since we have shown that the CNN is more successful at classification than the FFNN, we only consider the CNN for the present confusion scheme. The results of training on raw spin and vortex configurations are shown in Fig.~\ref{fig:fig4}. Learning on the raw spins results in a $\badW$ shape rather than the expected $\W$. As mentioned above, the finite-size XY model has a non-zero magnetization for $T<\TKT$ and this algorithm can easily classify any division $T^*<\TKT$ by a threshold magnetization. This supports our hypothesis from section~\ref{subsec:finite} that trained on raw spins, a CNN learns the magnetization.

	When trained on vortices, the expected $\W$ shape emerges, although it is skewed because we choose our training data from a non-symmetric region around $\TKT$. Despite having a powerful deep network, it is unable to learn any arbitrary partition and performs best near $\TKT$. This may be attributed to the fact that for low $T$, the vortex configurations are fundamentally similar; there are few vortices and they are logarithmically bound. This is in contrast with the raw spin configurations which may posses distinguishing features like the magnetization. Near $\TKT$, the network can distinguish the phases with high accuracy because of the true physical partition due to vortex unbinding. At high $T^*$ the vortex configurations look sufficiently random that the network again misclassifies for an arbitrary partition.

	We also observe significant finite-size effects in the $\W$ and $\badW$ shape, both broadening and shallowing with increasing $L$. The finite-size scaling behavior of the peak does not trend towards $\TKT$ in the vortex case, but rather always stays above it, similar to the specific heat peak (see Fig.~\ref{fig:4c}). Surprisingly, in Fig.~\ref{fig:4c}, we see the confusion scheme achieves higher accuracy at $T^*\approx 1$ than $\TKT=0.89$, which indicates that the
	false $T^*\approx1$ phase boundary is easier for the network to learn than the temperature $\tTKT$ predicted by the universal jump. While this effect might disappear in the thermodynamic limit, it is still troubling. Matters are even worse for training on raw spins since all $T^* < \TKT$ have accuracy greater than 98.5\% for $L=64$, so it is even unclear where $\tTKT$ is.

	The confusion scheme for the XY model offers insight into what our CNN prefers to learn. In the case of the raw spin configurations, we infer that it learns the finite magnetization of the spin configurations instead of topological features. Near $\TKT$, the network trained on vortices achieves slightly higher accuracy (see Fig.~\ref{fig:4c}); therefore, in this case, the network would benefit from learning vortices. Despite this argument, we stress that we have no strong evidence that our CNN is even capable of finding vortices. To address this, in the next section we propose a custom network designed for vortex detection and test if it works in practice.

\subsection{Custom architecture for learning vortices}

	In the previous sections, we compared networks trained on the raw spin configurations to those trained on vortex configurations which were constructed manually (i.e., feature-engineered). We now explore the possibly of a custom network architecture designed specifically for learning vortices as an intermediate representation, before performing classification. It is one of the remarkable features of deep neural networks that each layer may represent a new level of abstraction~\cite{schmidhuber_deep_2015,schmidhuber_deep_2015,guo_deep_2016,krizhevsky_imagenet_2017}. For example, in facial image recognition, the first convolution layer may extract edges, while the final layer encodes complex features such as facial expressions~\cite{zeiler_visualizing_2014}. We aim to design a network which may similarly be interpreted as representing vortices in an intermediate layer.

\begin{figure}
		\includegraphics[width = 0.9\columnwidth]{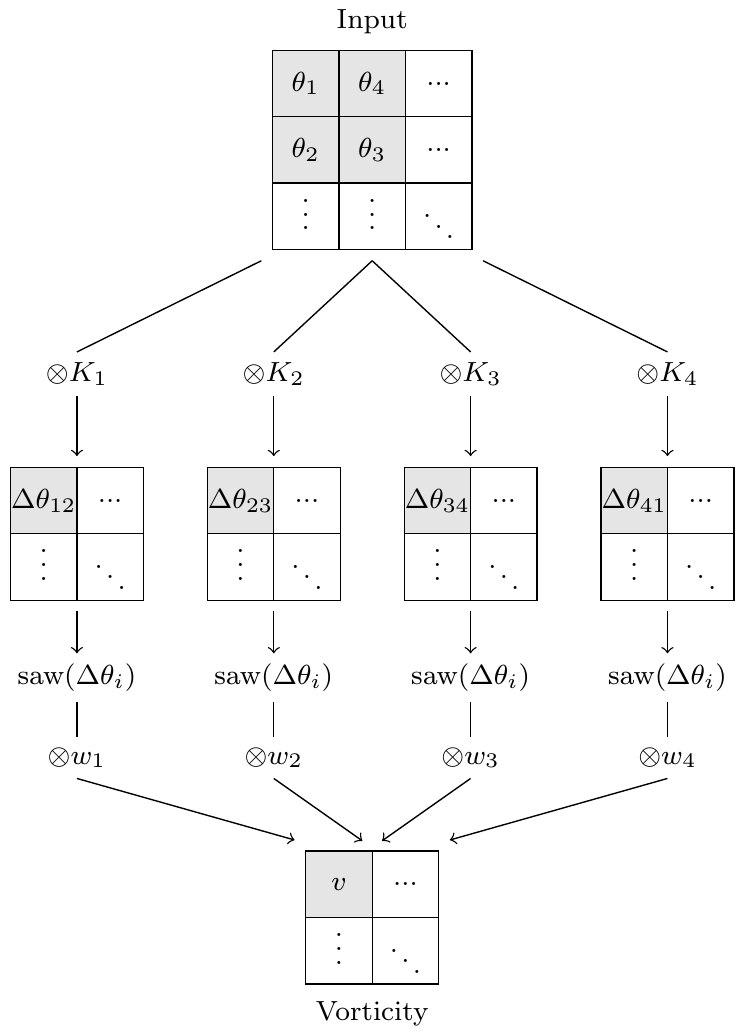}
		\caption{\label{fig:fig5} Visual representation of how the custom network architecture can compute the vorticity. We denote the convolution operation with $\otimes$, and ignore biases for the purpose of the diagram. Applying the four $2\times 2$ filters, $K_i$, partitions the data into four $L\times L$ arrays where each element is an angle differences in one lattice direction, $\Delta\theta_{ij}$. The angle differences are then converted into the range $\Delta\theta_{ij}\in[-\pi,\pi)$ by applying the sawtooth function from Eq.~(\ref{eq:saw}). A single $1 \times 1$ convolution filter with weights $w=[1,1,1,1]$ and zero biases then sums the four shifted angle differences into the vorticity.}
	\end{figure}

	Below, we derive the appropriate weights for a three-layer network which computes the vorticity from input spin configurations. The entire network is visualized in Fig.~\ref{fig:fig5}.
	The first layer, which acts on the input angle values, $\theta_i$, is a convolution layer with four $2\times 2$ convolution filters given by
	\begin{align}
		K_1 & = {\left[\begin{matrix}-1 & 1\\ 0 & 0\end{matrix}\right]}, & K_2 & = {\left[\begin{matrix}0 & -1\\ 0 & 1\end{matrix}\right]}, \\
		K_3 & = {\left[\begin{matrix}1 & 0\\ -1 &0 \end{matrix}\right]}, & K_4 & = {\left[\begin{matrix}0 & 0\\ 1 & -1\end{matrix}\right]}. \nonumber
	\end{align}
	The effect of these filters is to compute the nearest-neighbor angle differences, $\Delta \theta_{ij}$, within each plaquette.
	The next layer we apply is hard-coded to map the angle differences, $\Delta \theta_{ij}\in[-2\pi, 2\pi]$, into the range $[-\pi,\pi)$. This is done by applying the sawtooth function from Eq.~(\ref{eq:saw}) to each element in the $(L,L,4)$-dimensional array. The final processing layer computes a weighted sum of the four angle differences by applying a single $1 \times 1$ convolution filter. Uniform weights with zero biases would compute the vorticity exactly up to a multiplicative constant.
		
	While the network described above is capable of representing vortices within an internal layer (vorticity layer in Fig.~\ref{fig:fig5}), it might fail to do so in practice. To explore this we consider three possible variations of the initializations of the network parameters.

	The first variation consists of fixing the weights (and biases) in the first three layers such that the network computes the vorticity exactly. This is, of course, engineering the relevant features by hand; however, it provides a useful benchmark. The second variation is performed by initializing the weights exactly to those of the fixed network, then relaxing the constraints as training is continued. This step shows whether the original (vortex) minimum is stable. The third variation is simply the naive choice where the network parameters are initialized randomly.

	For all three variations, we train for binary classification by minimizing the cross-entropy loss from Eq.~(\ref{eq:loss}). Each network is trained 10 times with different validation sets. As per Section~\ref{subsec:finite}, we implement early stopping to terminate training once the loss function on the validation set fails to improve after 50 epochs. We train on lattice sizes from $L=8,...72$ in increments of 8.

\begin{figure}
		\includegraphics{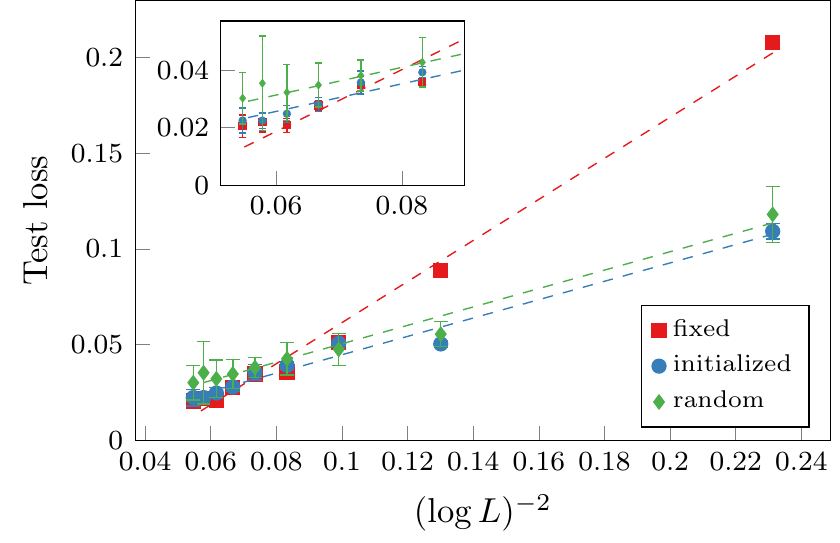}
		\caption{\label{fig:fig6}
		The loss function
		from Eq.~(\ref{eq:loss}) evaluated on the test set for three variations of the custom architecture for various lattice sizes $L$. For small $L<16$, the fixed network with hard-coded weights performs poorly compared to the others. For large $L>32$, the fixed network performs best, possibly due to a reduced number of trainable parameters. The inset shows a magnified region for $32\leq L \leq 72$.}
	\end{figure}

	We can understand the three variations by looking at the loss function evaluated on the test set as in Fig.~\ref{fig:fig6}. For small $L$, the loss function of the fixed network is much larger than the others, indicating that it is \emph{not} beneficial to represent the vortices for $L<16$. In this small-lattice region, learning vortices hinders classification. However, near $L\sim32$ the fixed network outperforms the other two. Hence, we conclude that only for the large-lattice region, $L>32$, is it beneficial for a network to learn an intermediate representation of the vorticity. This also agrees with the findings in Ref.~\cite{hu_discovering_2017} in which the topologically-invariant winding number could be learned for systems of size $L>32$.

\begin{figure}
	 	\sidesubfloat[]{\hspace{-.2cm}\includegraphics{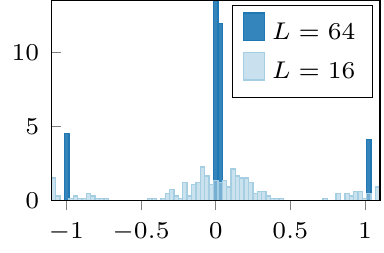}}\,
	 	\sidesubfloat[]{\hspace{-.15cm}\includegraphics{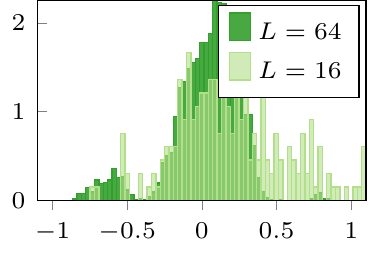}}
		\caption{\label{fig:hist} Histogram of the values of the vortex layer from Fig.~\ref{fig:fig5} which (ideally) computes the vorticity for: {\bf (a)} the network initialized to compute vorticity, and {\bf (b)} the randomly initialized network. In {\bf (a)}, we see for small $L$, the vorticity is not quantized, indicating that the network did not learn to compute the local vorticity; however, for large $L$, the histogram looks as expected for vortex detection. Conversely, the distribution in {\bf (b)} is appears unrelated to vortices for any $L$.}
	\end{figure}

	We can check what each network learns by looking at the histogram distribution of the outputs of the vorticity layer in Fig.~\ref{fig:fig5}. For the fixed network, we would see exactly integral quantities corresponding to the quantized vorticity. For the vortex-initialized network, Fig.~\ref{fig:hist} shows that for small $L$, it does not learn the true vorticity distribution, but for $L\geq32$ it does. This is consistent with the hypothesis that learning vortices is only beneficial for $L>32$. The randomly initialized network does not produce a histogram consistent with the learning of vortices for any system size studied.

	Interpreting the behavior of the neural network for large $L$ is not straightforward. As Fig.~\ref{fig:fig6} shows, the model with fixed features and less trainable parameters performs better for large $L$. This can likely be attributed to a lower-dimensional optimization landscape.
	We cannot conclude whether the vortex representation is a global minimum for the fully adjustable (randomly initialized) network variation. While it certainly performs best in fixed computational time, the higher dimensionality of the adjustable network may have an other global minimum not present in the lower-dimensional case. We can claim, however, that the vortex minimum is at least a stable local minimum since a network initialized to it never escapes, as demonstrated by the initialized variation for large $L$ in Fig.~\ref{fig:fig6}.

	Adding a custom regularization term could potentially alter the optimization landscape to aid the network in detecting vortices. One method would be to enforce integral quantities for an intermediate output of the network, but in our attempts, this results in the intermediate quantity peaked sharply around zero. There is also the possibility of adding a regularization to the initial kernels to learn only nearest-neighbors interactions, but this is overly restrictive and defeats the purpose of automated machine learning.

\section{Conclusion}

	In this paper, we asked whether it is possible for a neural network to learn the vortex unbinding at the KT transition in the two-dimensional classical XY model. We demonstrated the significant effects that feature engineering and finite lattice sizes have on the performance of supervised learning algorithms.

	Treating spin configurations as raw images and training on the thermodynamic value for $\TKT$, we found that naive supervised learning with a feed-forward network failed to converge to an accurate estimate for the KT transition temperature for moderate lattice sizes ($L\approx 32$). Conversely, a convolutional network performed consistently well with increasing $L$. Since the prediction of $\TKT$ from the convolutional network was insensitive to $L$, we inferred that the network extracted features related to the magnetization, which are present in any finite-size lattice. This conclusion was further supported by the observation that in the confusion method any false phase boundary $T^*$ below $\TKT$ could easily be learned by a network when trained on the raw spin configurations.

	By preprocessing the spin configurations into vorticity, both network architectures displayed finite-size scaling behavior consistent with the thermodynamic value of $\TKT$. In particular, the performance of the convolutional network continually improved as the system size increased, whereas the one-layer network's performance plateaued around $L=32$. When the confusion scheme was trained on vortices it did not predict the correct critical temperature; instead, the test accuracy reached a maximum near $T^* \approx 1$. This demonstrates the need for further study of the confusion scheme for the semi-supervised learning of phase transitions.

	We further explored if such extreme feature engineering could be relaxed while retaining acceptable accuracy. We devised a deep-layered structure of weights that could be constrained to extract vortices from the raw spin configurations or left free to explore other minima in the learning process. We found that it is beneficial for the network to discover vortices only for lattices with of over 1000 spins. Yet, even for large system sizes, a randomly-initialized network settled into a local minimum not related to vortices. It is likely that the optimization landscape of our designed network is sufficiently rough so that stochastic gradient descent would take exponentially long to find a minimum where the learned features correspond to vortices.

	The difficultly that these standard supervised learning techniques have in discriminating the phases of the XY model underscores the challenge that unsupervised learning techniques could face in learning the KT transition from unlabeled data. Our work emphasizes the need for further study into how much feature engineering is required before topological features can be used reliably for the machine learning of unconventional phases and phase transitions.

\begin{acknowledgments}
	The authors would like to thank
	J.~Carrasquilla,
	C.~X.~Cerkauskas,
	L.~E.~Hayward Sierens,
	B.~Kulchytskyy,
	R.~Lewitzky,
	A.~Morningstar,
	P.~Ponte,
	J.~Rau,
	S.~Sachdev,
	R.~Scalettar,
	R.~Singh,
	G.~Torlai,
	F.~Verstraete,
	and C.~Wang
	for many useful discussions.
 	This research was supported by the Natural Sciences and Engineering Research Council of Canada (NSERC), the Canada Research Chair program, the Perimeter Institute for Theoretical Physics, and the National Science Foundation under Grant No. NSF PHY-1125915. This work was made possible by the facilities of the Shared Hierarchical Academic Research Computing Network (SHARCNET) and Compute/Calcul Canada. We also gratefully acknowledge the support of NVIDIA Corporation with the donation of the Titan Xp GPU used for this research. Research at Perimeter Institute is supported by the Government of Canada through Industry Canada and by the Province of Ontario through the Ministry of Research \& Innovation.
\end{acknowledgments}

\bibliography{ml-xy-paper.bbl}

\end{document}